\documentclass[aps, superscriptaddress, twocolumn]{revtex4-1}

\usepackage[utf8]{inputenc}
\usepackage{lmodern}
\usepackage[version=3]{mhchem}
\usepackage{amsmath,amssymb,amstext,amsfonts}
\usepackage{natbib}
\usepackage{psfrag,graphicx}
\usepackage[]{units}
\usepackage{upgreek}
\usepackage{textcomp} 
\usepackage{color}
\usepackage{xcolor}
\usepackage{float}
\usepackage{array} 
\usepackage{multirow}
\usepackage{epstopdf}
\usepackage{graphicx}
\usepackage{dcolumn}
\usepackage{bm}

\usepackage[utf8]{inputenc}
\usepackage[T1]{fontenc}
\usepackage{mathptmx}
\usepackage[nolist]{acronym}
\usepackage{subcaption}
\usepackage{siunitx}
\usepackage[hidelinks]{hyperref}
\usepackage{lipsum}

\DeclareSIUnit{\wtpercent}{wt\%}
\DeclareSIUnit{\volpercent}{vol\%}

\usepackage{pgf}
\usepackage{tikz}
\usepackage{pgfplots}
\usetikzlibrary{
	matrix,
	pgfplots.external,
	pgfplots.groupplots,
	positioning,
	pgfplots.colormaps,
	colorbrewer
}
\pgfplotsset{width=5cm,compat=newest}
\usetikzlibrary{colorbrewer}
\usepgfplotslibrary{colorbrewer}
\pgfplotsset{cycle list/Set1}

\usetikzlibrary{pgfplots.colormaps}

\definecolor{01}{HTML}{1F77B4}
\definecolor{02}{HTML}{AEC7E8}
\definecolor{03}{HTML}{FF7F0E}
\definecolor{04}{HTML}{FFBB78}
\definecolor{05}{HTML}{2ca02c}
\definecolor{06}{HTML}{d62728}

\begin{document}


\title{3D printing of polymer-bonded anisotropic magnets in an external magnetic field and by a modified production process}

\author{Klaus Sonnleitner}
 \email{klaus.sonnleitner@univie.ac.at}
 
\author{Christian Huber}
\affiliation{ 
Physics of Functional Materials, University of Vienna, 1090 Vienna, Austria
}
\affiliation{ 
Christian Doppler Laboratory for Advanced Magnetic Sensing and Materials, 1090 Vienna, Austria
}
\author{Iulian Teliban}
\affiliation{ 
Magnetfabrik Bonn GmbH, 53119 Bonn, Germany
}

\author{Spomenka Kobe}
\affiliation{ 
Department of Nanostructured Materials, Jožef Stefan Institute, 1000 Ljubljana, Slovenia
}

\author{Boris Saje}
\affiliation{ 
Kolektor Magnet Technology GmbH, 45356 Essen, Germany
}

\author{Daniel Kagerbauer}
\affiliation{ 
Atominstitut, TU Wien, 1020 Vienna, Austria
}

\author{Michael Reissner}
\affiliation{ 
Low Temperature Facilities, TU Wien, 1040 Vienna, Austria
}

\author{Christian Lengauer}
\affiliation{ 
Department of Mineralogy and Crystallography, University of Vienna, 1090 Vienna, Austria
}

\author{Martin Groenefeld}
\affiliation{ 
Magnetfabrik Bonn GmbH, 53119 Bonn, Germany
}

\author{Dieter Suess}
\affiliation{ 
Physics of Functional Materials, University of Vienna, 1090 Vienna, Austria
}
\affiliation{ 
Christian Doppler Laboratory for Advanced Magnetic Sensing and Materials, 1090 Vienna, Austria
}

\date{\today}

\begin{abstract}
The possibility of producing polymer-bonded magnets with the aid of additive processes, such as 3D printing, opens up a multitude of new areas of application. 
Almost any structures and prototypes can be produced cost-effectively in small quantities.
Extending the 3D printing process allows the manufacturing of anisotropic magnetic structures by aligning the magnetic easy axis of ferromagnetic particles inside a paste-like compound material along an external magnetic field. This is achieved by two different approaches: First, the magnetic field for aligning the particles is provided by a permanent magnet. Secondly, the 3D printing process itselfs generates an anisotropic behavior of the structures. An inexpensive and customizable end-user fused filament fabrication 3D printer is used to print the magnetic samples. The magnetical properties of different magnetic anisotropic Sr ferrite and SmFeN materials are investigated and discussed.
\end{abstract}

\maketitle

\ac{FFF} is a well established additive manufacturing (3D printing) process using thermoplastic materials \cite{zein2002fused,tofail2018additive}. Key advantages of \ac{FFF} are: It is an affordable, fast and end-user friendly manufacturing process. This work deals with the 3D printing of polymer-bonded and magnetic anisotropic magnets. Polymer-bonded magnets are composed of magnetic powder and a polymer matrix. Therefore, the maximum energy product $(BH)_\textrm{max}$ of bonded magnets is limited compared to sintered magnets. Nevertheless, many applications need an accurate and complex magnetic field distribution instead of highest field strengths, for example in sensor and electric drive technology \cite{elian2014integration}. Polymer-bonded magnets can be produced from magnetic powder that is magnetically isotropic or anisotropic. If a high energy product $(BH)_\textrm{max}$ of polymer-bonded magnets is not the most important parameter, magnetic isotropic powder is preferred because it is associated with lower costs and more flexibility. On the other hand, anisotropic magnets can only be magnetized in one certain direction, which can be restrictive with respect to shape. For producing polymer-bonded magnets, magnetic powder is mixed with a binder such as a thermoplastic polymer. The resulted compound can be used for injection molding, compression and for extrusion. For the FFF process, thermoplastic or compounds  are extruded into wire-shaped filaments. Due to the high filler content of the compound, the viscosity increases which can lead to filling and flowing problems. Polyamides such as PA6, PA11 and PA12 offer a good combination of viscosity, bonding and mechanical properties such as a high tensile- and impact strength \cite{xiao2000modeling,salmoria2012mechanical}. For 3D printing, PA6 is more difficult to process because the water absorption capacity is significantly higher compared to PA12: PA6 absorbs approximately \SI{1.13}{\wtpercent}, while PA12 absorbs \SI{0.15}{\wtpercent} \cite{rajesh2002effect}. To prevent printing errors, PA6 compounds have to be dried at around \SI{90}{\celsius} for \SI{8}{\hour}.

Certain applications of magnetic designs have complex requirements on the magnetic field distribution, for example highest possible magnetic field homogeneity in a certain volume. 3D printing of isotropic NdFeB magnets with focus on tailoring the external stray field by topology optimisation of the magnet was successfully implemented. \cite{huber20173d,huber20163d,huber2017topology}. With \ac{BAAM} isotropic NdFeB magnets were manufactured \cite{li2016big}, where a special heat and extruding unit is needed and only large structures can be realized. Goal of this work is to produce magnetic anisotropic polymer-bonded magnets with the help of two different approaches by an end user 3D printer: Printing directly on the surface of a strong permanent magnet and investigating how the printing process itself affects the magnetic anisotropy of the sample.

Three different magnetic anisotropic compounds are investigated:
\begin{enumerate}
	\item  Strontium Hexaferrite inside a PA6 matrix (Sprox\textsuperscript{\circledR}10/20p), fill grade: \SI{49}{\volpercent},
	\item  Strontium Hexaferrite inside a PA12 matrix (Sprox\textsuperscript{\circledR}11/22p), fill grade: \SI{53}{\volpercent},
	\item $\mathrm{Sm_{2}Fe_{17}N_{3}}$ inside a PA12 matrix, fill grade: \SI{44}{\volpercent}.
\end{enumerate}
The two Sprox\textsuperscript{\circledR} compounds are prefabricated by the Magnetfabrik Bonn. $\mathrm{Sm_{2}Fe_{17}N_{3}}$ and the thermoplastic PA12 are compounded with twin screw extruder at the University of Leoben, Austria\cite{kracalik2009advanced}. The advantage of these ferrite powders is that they are inexpensive but have a lower $(BH)_\textrm{max}$ compared to rare earth materials. 
The volume filler constant for these materials is determined by the \ac{LOI}, whereby the sample is heated to \SI{1100}{\celsius} and the plastic is evaporated\cite{heiri2001loss}.

\begin{figure}[htbp]
	\includegraphics[scale=1.0]{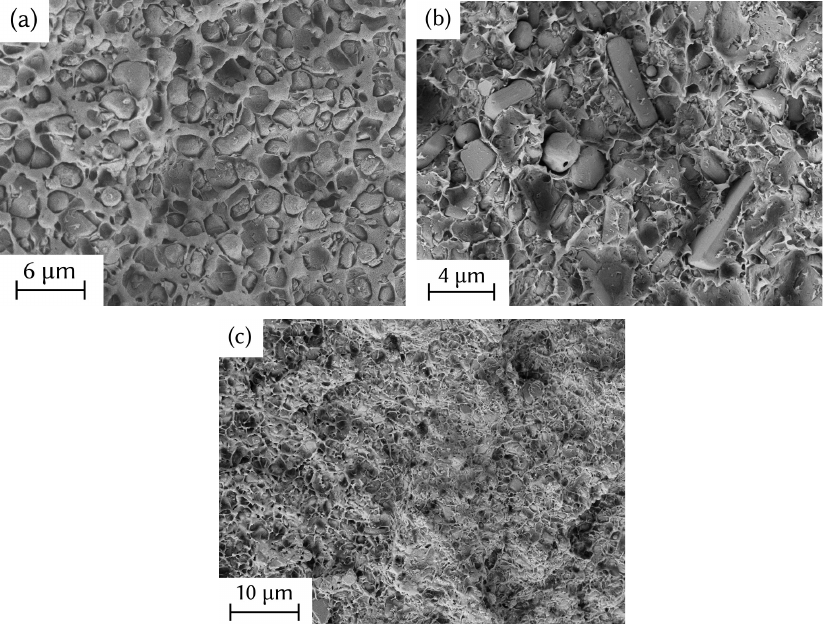}
	\caption{Images of a \acl{SEM}: (a) $\mathrm{Sm_{2}Fe_{17}N_{3}+PA12}$, (b) Sprox\textsuperscript{\circledR}10/20p and (c) Sprox\textsuperscript{\circledR}11/22p.}
	\label{fig:SEM}
\end{figure}

Pictures generated with a \ac{SEM} reveal information of the microstructure of the materials. The Strontium Hexaferrite grains are platelets with \SI[product-units = power]{6x2x2}{\micro\metre}~$(L \times W \times H)$ and with the basic magnetic properties of $B_\text{r}=\SI{196}{mT}$ and with $H_\text{cJ}=\SI{183}{kA\per m}$, Fig. \ref{fig:SEM}\,(b) and (c).
$\mathrm{Sm_{2}Fe_{17}N_{3}}$ particles are approximately spherical with a diameter of \SIrange[range-units = single]{3}{4}{\micro\metre} with $B_\text{r}=\SI{1.31}{T}$ and $H_\text{cJ}=\SI{889}{kA\per m}$, Fig. \ref{fig:SEM}\,(a). 


By applying an external magnetic field with varying strength during the printing process, the minimum alignment field to orient the magnetic particles can be determined. The Sprox\textsuperscript{\circledR} and $\mathrm{Sm_{2}Fe_{17}N_{3}}$ grains are particles with uniaxial anisotropy and show one magnetic easy axis. In our case, we are using a permanent magnet to generate an external field during the printing process. The used magnets provide fields of approximately \SI{550}{mT} for a \SI[product-units = power]{40x40x40}{\milli\metre}~$(L \times W \times H)$ NdFeB magnet in the center and \SI{150}{mT} for a plate cylinder with a diameter of \SI{25}{mm} and a thickness of \SI{3}{mm}. The calculation of the flux density along the symmetry axis of a cube and a cylinder magnet is performed analytically \cite{camacho2013alternative}. The distance between the permanent magnet and the printed sample is varied. The distances are chosen to create a $B_\mathrm{z}$ of \SIlist[list-units = single]{200;150;100}{\milli\tesla}. As the thermoplastic matrix melts, the particles can orient itself under the external field to form an anisotropic magnet. It is necessary to print the magnets as close as possible to the center of the permanent magnet since the field gradients increase with increasing distance from the symmetry axis. As a consequence a force acts on the magnetic particle which leads to deflection and a unfavorable magnetization of the sample. The advantage of this method is that the printer does not require any modifications, only the permanent magnet has to be attached to the printing bed.
According to the Stoner-Wohlfarth model the remanent magnetization $m_\mathrm{r}$ and the coercivity $h_\mathrm{c}$ are reduced by a factor of $0.5$ for an assembly of randomly oriented particles in comparison to a fully aligned structure \cite{tannous2008stoner}. This means  $m_\mathrm{r}/m_\mathrm{s}=0.5$ for an assembly of not aligned, single domain Stoner-Wohlfarth particles and $m_\mathrm{r}/m_\mathrm{s}=1$ for an assembly of parallel to the easy axis aligned, single domain particles, where $m_\mathrm{s}$ is the saturation magnetization. The hysteresis loop is measured in the \ac{VSM} and reveals how good the particles in the matrix are aligned. The samples examined are cubes and are printed with a modified \ac{FFF} printer. Cubes with the size \SI[product-units = power]{10x10x5}{\milli\metre}~$(L \times W \times H)$ are printed and then processed to a smaller format: \SI[product-units = power]{3x3x3}{\milli\metre}~$(L \times W \times H)$. This was necessary because printing the sample directly is difficult and the \ac{VSM} supports only geometries smaller than \SI[product-units = power]{4x4x4}{\milli\metre}~$(L \times W \times H)$ where the dipole approximation is still valid. The used printer, Velleman's K8200, allows the exchange of components, since the printer software Marlin is open source \cite{moore2016vulnerability} and temperatures up to \SI{350}{\celsius} can be reached by replacing the stock extruder with E3D's Titan Aero. The viscosity of the material can be reduced by increasing the printing temperatures, which has a positive effect on processing. The optimal print parameters for the three materials are shown in Tab. \ref{tab:printParameter}.

\begin{table}[htbp]
	\caption{\label{tab:printParameter} Best parameter settings for printer and Slic3r\cite{ranellucci2013reprap}, empirically found.}
	\begin{ruledtabular}
		\begin{tabular}{lr}
			Parameter & Value\\
			\hline
			Extruder temp & \SI{300}{\celsius} \\
			Layer height & \SI{0.25}{\milli\metre}  \\
			Fill density & \SI{100}{\percent} \\
			Fill pattern & rectilinear \\
			Printer speed & \SIrange[range-units = single]{15}{20}{\milli\metre\per\second} \\
			Build platform & painter's tape \\
			Bed temperature & \SI{40}{\celsius} \\
		\end{tabular}
	\end{ruledtabular}
\end{table}

\begin{figure}[htbp]
	\includegraphics[scale=1.0]{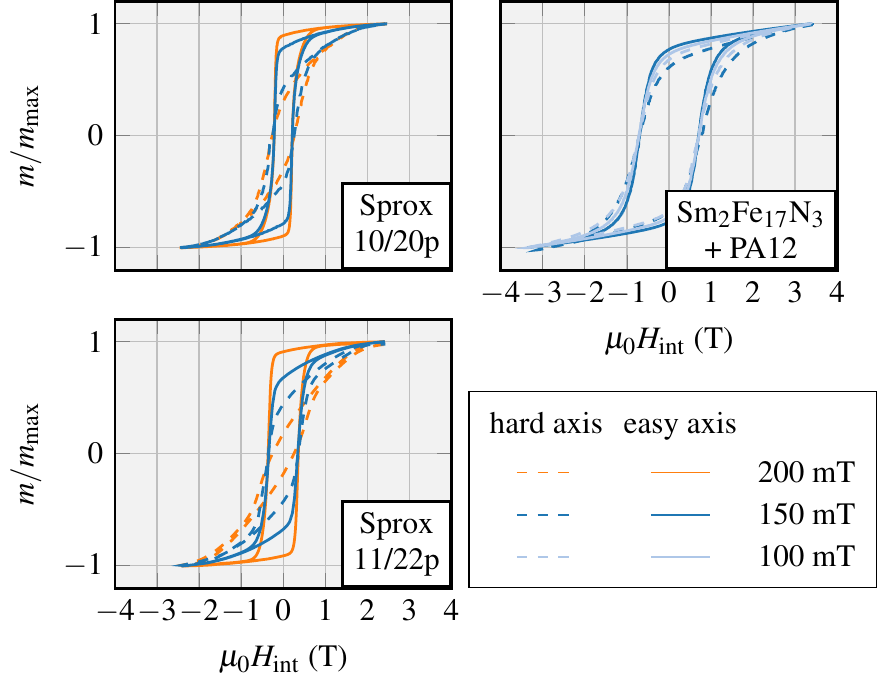}
	\caption{Hysteresis loops in hard- and easy axis of three samples measured with the \ac{VSM}, where $H_\mathrm{int}$ is the internal field, considering a demagnetization factor of $N=1/3$. The samples are aligned with different magnitudes of the external field.}
	\label{fig:VSM_print_on_cube}
\end{figure}

Fig.\,\ref{fig:VSM_print_on_cube} shows the hysteresis loops of all three materials in hard and easy axis, with samples printed at different external magnetic flux densities $\mu_0 H_\mathrm{ext}=$~\SIlist[list-units = single]{100;150;200}{\milli\tesla} and Tab.\,\ref{tab:comparison_mr_ms} lists the corresponding ratio $m_\mathrm{r}/m_\mathrm{s}$ with the remanence $m_\mathrm{r}$ and the saturation magnetization $m_\mathrm{s}$. The internal field $\mu_0 H_\mathrm{int} = \mu_0 H_\mathrm{ext} - N m$ is displayed on the $x$-axis, assuming a demagnetization factor of $N=1/3$ and $m$ is the magnetization. This applies to a perfect cube and for real cubes small deviations are to be expected \cite{aharoni1998demagnetizing}.

\begin{table}[htbp]
	\centering
	\caption{The ratio $m_\mathrm{r}/m_\mathrm{s}$ in magnetic easy axis of samples printed at various magnetic flux densities.}
	\begin{tabular}{| c | c | c | c |}
		\hline
		$m_\mathrm{r} / m_\mathrm{s}$ & 200 mT &  150 mT &  100 mT \\ \hline
		Sprox 10/20p & 0.88 & 0.76 & - \\
		Sprox 11/22p & 0.94 & 0.42 & 0.4 \\
		$\mathrm{Sm_{2}Fe_{17}N_{3}}$+PA12 & - & 0.75 & 0.7 \\
		\hline
	\end{tabular}
	\label{tab:comparison_mr_ms}
\end{table}

For both Sprox\textsuperscript{\circledR} materials an external flux density of \SI{200}{\milli\tesla} is sufficient for aligning the particles because of their lower  coercive field. However, \SI{150}{\milli\tesla} is the highest external field in which $\mathrm{Sm_{2}Fe_{17}N_{3}+PA12}$ can be printed, which is not enough for alignment. Larger flux densities destroy the shape of the sample due to the larger remanence magnetization of the material compared to Sprox\textsuperscript{\circledR}. Also the density and therefore the remanence of the printed sample is reduced compared to the raw compound, Tab. \ref{tab:density_and_remanence_printed_sample_vs_compound}. All values in the table, measured and from the data sheet, refer to anisotropic samples.

\begin{table}[htbp]
	\centering
	\caption{$B_\mathrm{r1}$, $H_\mathrm{cJ1}$ and $\rho_\mathrm{1}$ measured properties of printed anisotropic sample, $B_\mathrm{r2}$, $H_\mathrm{cJ2}$ and $\rho_\mathrm{2}$ according to datasheet. For $\mathrm{Sm_{2}Fe_{17}N_{3}+PA12}$ the printed anisotropic ($B_\mathrm{r1}$) and printed isotropic ($B_\mathrm{r2}$) samples are compared.}
	\begin{tabular}{| c | c | c | c |}
		\hline
		 & Sprox 10/20p &  Sprox 11/22p &  $\mathrm{Sm_{2}Fe_{17}N_{3}+PA12}$ \\ \hline
		$B_\mathrm{r1}$ (\si{\milli\tesla})& 201 & 220 & 308 \\
		$H_\mathrm{cJ1}$ (\si{kA\per m})& 162 & 281 & 565 \\
		$B_\mathrm{r2}$ (\si{\milli\tesla})& 222 & 225 & 389\footnote{calculated} \\
		$H_\mathrm{cJ2}$ (\si{kA\per m})& 207 & 239 & 899 \\
		$\rho_\mathrm{1}$ (\si{\gram\per\cubic\centi\metre})& 2.861 & 2.962 & 3.404 \\
		$\rho_\mathrm{2}$ (\si{\gram\per\cubic\centi\metre}) & 3.2 & 3.2 & 3.796\footnote{density measurement of compound, no data sheet available} \\
		$B_\mathrm{r1}/B_\mathrm{r2}$ & 0.91 & 0.98 & 0.79 \\
		$\rho_\mathrm{r1}/\rho_\mathrm{r2}$ & 0.89 & 0.91 & 0.9 \\
		\hline
	\end{tabular}
	\label{tab:density_and_remanence_printed_sample_vs_compound}
\end{table}

\begin{figure}[htbp]
	\includegraphics[scale=1.0]{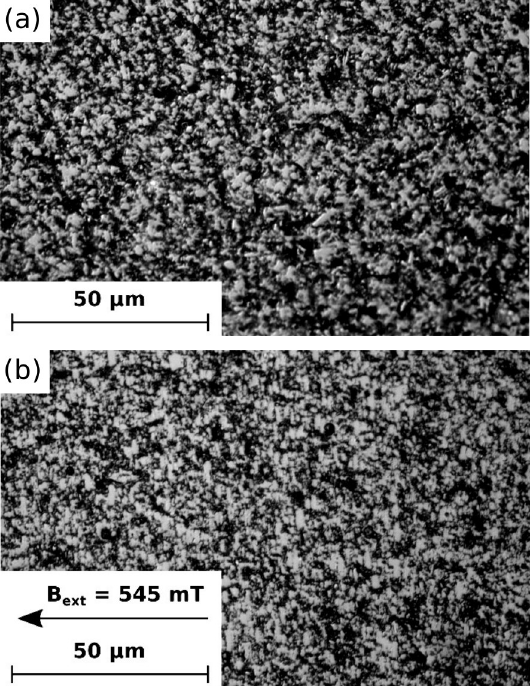}
	\caption{Reflected light microscope images: (a) Isotropic Sprox\textsuperscript{\circledR}11/22p (b) aligned anisotropic particles of the printed magnet in an external magnetic field of \SI{545}{\milli\tesla}.}
	\label{fig:Zeiss_Axioplan_550mT_and_Sprox}
\end{figure}

A further functional principle is being investigated in which the printing direction is observed in relation to the orientation of the particles, the so-called "flow anisotropy". The magnetic hard and easy axis depend on the geometry: The magnetic easy axis of the Sprox\textsuperscript{\circledR} particles is perpendicular to the long side of their cuboid geometry. The orientation of the particles is shown with a light microscope in reflected light mode. Fig.~\ref{fig:Zeiss_Axioplan_550mT_and_Sprox}\,(a) shows an isotropic sample and Fig.~\ref{fig:Zeiss_Axioplan_550mT_and_Sprox}\,(b) a sample that is printed under an external field of \SI{545}{\milli\tesla}. By comparing the images in Fig.~\ref{fig:Zeiss_Axioplan_550mT_and_Sprox} one can see the alignment of the particles in the external magnetic field in vertical direction.

\begin{figure}[htbp]
	\includegraphics[scale=1.0]{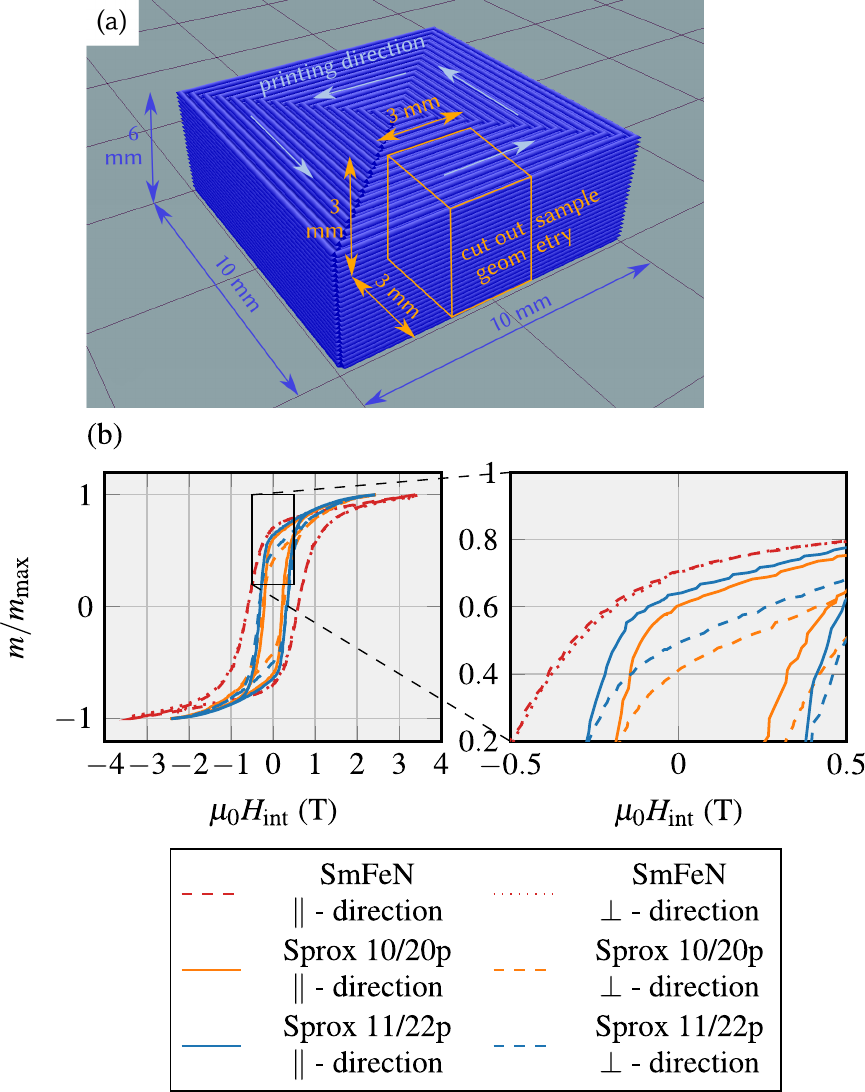}
	\caption{(a) Every vertical shell of the dark blue sample (\SI[product-units = power]{10x10x6}{\milli\metre}~$(L \times W \times H)$) is printed in direction of the light blue arrows consisting only of perimeters (walls). The sample for the \ac{VSM} measurement (orange) is cut out of the dark blue sample: all layers including infill and thus the whole volume (\SI[product-units = power]{3x3x3}{\milli\metre}~$(L \times W \times H)$) of the resulting cube are printed in the same direction. (b) Hysteresis loops in direction of printing and orthogonal to print direction of three samples measured with the \ac{VSM}, where $H_\mathrm{int}$ is the internal field.}
	\label{fig:flowanisotropy_sample_and_VSM}
\end{figure}

The cuboid structure of Sprox\textsuperscript{\circledR} allows mechanical orientation over the printing direction whereas the spherical structure of SmFeN remains unaffected. Fig.~\ref{fig:flowanisotropy_sample_and_VSM}\,(a) shows how this idea is realized: When a cuboid is printed consisting only of perimeters, areas arise where print direction is the same over the whole volume of the sample. For soft magnetic materials, the influence of the 3D printing process was investigated by Patton et al. \cite{patton2019manipulating}.

Hysteresis loops with the applied field parallel to and orthogonal to the print direction were performed in the \ac{VSM} with the \SI[product-units = power]{3x3x3}{\milli\metre}~$(L \times W \times H)$ cubes. Fig.~\ref{fig:flowanisotropy_sample_and_VSM}\,(b) shows the hysteresis loops of the different samples. The SmFeN sample shows no difference between the two measured directions, whereas for the Sprox samples a difference is visible. This is a result of the Sprox aligned in the printing direction and no alignement of the SmFeN sample.
The values of the normalized remanance $m_\mathrm{r} / m_\mathrm{s}$ are shown in Tab.~\ref{tab:flowaniso_comparison_mr_ms}.

\begin{table}[htbp]
	\centering
	\caption{The ratio $m_\mathrm{r}/m_\mathrm{s}$ in parallel and perpendicular direction of print - movement.}
	\begin{tabular}{| c | c | c |}
		\hline
		$m_\mathrm{r} / m_\mathrm{s}$ & $\parallel$ - direction & $\perp$ -direction \\ \hline
		Sprox 10/20p & 0.58 & 0.39 \\
		Sprox 11/22p & 0.62 & 0.4 \\
		SmFeN & 0.66 & 0.67  \\
		\hline
	\end{tabular}
	\label{tab:flowaniso_comparison_mr_ms}
\end{table}

This paper presents two methods for aligning the easy axis of magnetic particles during 3D printing. A ferromagnetic powder inside a thermoplastic matrix is processed. The described techniques allow the printing of samples with a higher remanence $B_\mathrm{r}$ than isotropic powders. Compared to the presented approaches, providing an external alignment field produces better results in terms of $m_\mathrm{r}/m_\mathrm{s}$ but the observed flow anisotropy shows, that the print process itself has an influence on the alignment of the particles. The results show that a magnetic field $B_\mathrm{z}$ of approximately \SI{200}{\milli\tesla} is sufficient for aligning the the hardmagnetic Strontium Hexaferrite particles in the Sprox\textsuperscript{\circledR} compound materials. For SmFeN the threshold value stays unknown, since printing above external fields bigger than \SI{150}{\milli\tesla} is not possible right now. An improved cooling could contribute to the increase of the magnetic field during printing SmFeN. As a next step the results motivate the development of a customized 3D printer capable of aligning the particles arbitrarily during printing by applying a variable external magnetic field of \SI{200}{\milli\tesla} in the desired direction for the alignment of the particle. After successfully implementing this design, complex structures with special magnetic capabilities should be printable. This would be a breakthrough in the development and manufacturing of magnets with varying local magnetization direction that can not be produced by fabrication techniques that are state of the art. 

The support of the CD-Labor AMSEN (financed by the Federal Ministry of Economics, Family and Youth, the National Foundation for Research, Technology and Development) is acknowledged. The \ac{SEM} sample preparation and image generation is performed at the Faculty Center for Nano Structure Research at the University of Vienna, Austria.


%

\begin{acronym}[slmtA]
	\acro{VVSM}{Vector Vibrating Sample Magnetometer}
	\acro{VSM}{Vibrating Sample Magnetometer}
	\acro{SLS}{Selective Laser Sintering}
	\acro{FFF}{Fused Deposition Modeling}
	\acro{SLA}{Stereolithograhpy}
	\acro{AM}{Additive Manufacturing}
	\acro{CAD}{Computer-Aided Design}
	\acro{FEM}{finite element}
	\acro{PLA}{Polylactide}
	\acro{MA}{Magnetic Anisotropy}
	\acro{SEM}{Scanning Electron Microscope}
	\acro{FFF}{Fused Filament Fabrication}
	\acro{LOI}{Loss on Ignition}
	\acro{BAAM}{Big Area Additive Manufacturing}
	
	
	
\end{acronym}

\end{document}